# Guidelines for including grey literature and conducting multivocal literature reviews in software engineering

Vahid Garousi
Information Technology Group
Wageningen University, Netherlands
vahid.garousi@wur.nl

Michael Felderer
University of Innsbruck, Austria &
Blekinge Institute of Technology, Sweden
michael.felderer@uibk.ac.at

Mika V. Mäntylä
M3S, Faculty of Information Technology
and Electrical Engineering
University of Oulu, Oulu, Finland
mika.mantyla@oulu.fi

**Abstract:**

*Context*: A Multivocal Literature Review (MLR) is a form of a Systematic Literature Review (SLR) which includes the grey literature (e.g., blog posts, videos and white papers) in addition to the published (formal) literature (e.g., journal and conference papers). MLRs are useful for both researchers and practitioners since they provide summaries both the state-of-the art and –practice in a given area. MLRs are popular in other fields and have recently started to appear in software engineering (SE). As more MLR studies are conducted and reported, it is important to have a set of guidelines to ensure high quality of MLR processes and their results.

*Objective*: There are several guidelines to conduct SLR studies in SE. However, several phases of MLRs differ from those of traditional SLRs, for instance with respect to the search process and source quality assessment. Therefore, SLR guidelines are only partially useful for conducting MLR studies. Our goal in this paper is to present guidelines on how to conduct MLR studies in SE.

*Method*: To develop the MLR guidelines, we benefit from several inputs: (1) existing SLR guidelines in SE, (2), a literature survey of MLR guidelines and experience papers in other fields, and (3) our own experiences in conducting several MLRs in SE. We took the popular SLR guidelines of Kitchenham and Charters as the baseline and extended/adopted them to conduct MLR studies in SE. All derived guidelines are discussed in the context of an already-published MLR in SE as the running example.

*Results*: The resulting guidelines cover all phases of conducting and reporting MLRs in SE from the planning phase, over conducting the review to the final reporting of the review. In particular, we believe that incorporating and adopting a vast set of experience-based recommendations from MLR guidelines and experience papers in other fields have enabled us to propose a set of guidelines with solid foundations.

*Conclusion*: Having been developed on the basis of several types of experience and evidence, the provided MLR guidelines will support researchers to effectively and efficiently conduct new MLRs in any area of SE. The authors recommend the researchers to utilize these guidelines in their MLR studies and then share their lessons learned and experiences.

## TABLE OF CONTENTS

## 1 INTRODUCTION

Systematic Literature Reviews (SLR) and Systematic Mapping (SM) studies were adopted from medical sciences in mid-2000's [1], and since then numerous SLRs studies have been published in software engineering (SE) [2, 3]. SLRs are valuable as they help practitioners and researchers by indexing evidence and gaps of a particular research area, which may consist of several hundreds of papers [4-9]. Unfortunately, SLRs fall short in providing full benefits since they typically review the formally-published literature only while excluding the large bodies of the "grey" literature (GL), which are constantly produced by SE practitioners outside of academic forums [10]. As SE is a practitioner-oriented and an application-oriented field [11] the role of GL should be formally recognized, as has been done for example in educational research [12, 13] and health sciences [14-16], and management [17]. We think that GL can enable a rigorous identification of emerging research topics in SE as many research topics already stem from software industry.

SLRs which include both the academic and the GL were termed as Multivocal Literature Reviews (MLR) in educational research [12, 13], in the early 1990's. The main difference between an MLR and an SLR is the fact that, while SLRs use as input only academic peer-reviewed papers, MLRs in addition also use sources from the GL, e.g., blogs, videos, white papers and web-pages [18]. MLRs recognize the need for "multiple" voices rather than constructing evidence from only the knowledge rigorously reported in academic settings (formal literature). The MLR definition from [12] elaborates this: "*Multivocal literatures are comprised of all accessible writings on a common, often contemporary topic. The writings embody the views or voices of diverse sets of authors (academics, practitioners, journalists, policy centers, state offices of education, local school districts,* *independent research and development firms, and others). The writings appear in a variety of forms. They reflect different purposes, perspectives, and information bases. They address different aspects of the topic and incorporate different research or non-research logics*".

Many SLR recommendations and guidelines, e.g., Cochrane [19], do not prevent including GL in SLR studies, but on the contrary, they recommend considering the GL as long as GL sources meet the inclusion/exclusion criteria [20]. Yet, nearly all SLR papers in the SE domain exclude GL in SLR studies, a situation which hurts both academia and industry in our field. To facilitate adoption of the guidelines we integrate boxes throughout the paper that cover concrete guidelines summarizing more detailed discussions of specific issues in the respective sections.

The purpose of this paper is therefore to promote the role of GL in SE and to provide specific guidelines for including GL and conducting multivocal literature reviews. We aim at complementing the existing guidelines for SLR studies [3, 21, 22] in SE to address peculiarities of including the GL in our field. Without proper guidelines, conducting MLRs by different teams of researchers may result in review papers with different styles and depth. We support the idea that, "*more specific guidelines for scholars on including grey literature in reviews are important as the practice of systematic review in our field continues to mature*", which originates from the field of management sciences [17]. Although multiple MLR guidelines have appeared in areas outside SE, e.g. [19, 20], we think they are not directly applicable for two reasons. First, the specific nature of GL in SE needs to be considered (the type of blogs, questions answer sites, and other GL sources in SE). Second, the guidelines are scattered to different disciplines and offer conflicting suggestions. Thus, in this paper we integrate them all and utilize our prior MLR expertise to present a single "synthesized" guideline.

This paper is structured similar to SLR [22] and SM guideline [3] in SE and considers three phases: (1) planning the review, (2) conducting the review, and (3) reporting the review results. The remainder of this guidelines paper is structured as follows. Section 2 provides a background on concepts of GL and MLRs. Section 3 explains how we developed the guidelines. Section 4 presents guidelines on planning an MLR, Section 5 on conducting an MLR, and Section 6 on reporting an MLR. Finally, in Section 8, we draw conclusions and suggest areas for further work.

## 2 BACKGROUND

We review the concept of GL in Section 2.1. We then discuss different types of secondary studies (of which MLR is a type of) in Section 2.2. Section 2.3 reviews the emergence of and need for MLRs in SE. We then motivate the need for a set of guidelines for conducting MLR studies in Section 2.4.

### 2.1 An overview of the concept of grey literature

We found several definitions of GL in the literature. The most widely used and accepted definition is the so-called Luxembourg definition which states that, "<*grey literature*> *is produced on all levels of government, academics, business and industry in print and electronic formats, but which is not controlled by commercial publishers, i.e., where publishing is not the primary activity of the producing body*" [23]. The Cochrane handbook for systematic reviews of interventions [24] defines GL as "*literature that is not formally published in sources such as books or journal articles*". Additionally, there is an annual conference on the topic of GL (www.textrelease.com) and an international journal on the topic (www.emeraldinsight.com/toc/ijgl/1/4). There is also a Grey Literature Network Service (www.greynet.org) which is "*dedicated to research, publication, open access, education, and public awareness to grey literature*".

To classify different types of sources in the GL we adopted an existing model from the management domain [17] to SE in Figure 1. The changes that we made to the model in [17] to make it more applicable to SE was a revision of the outlets on the right-hand side under the three "tier" categories, e.g., we added the Q/A websites (such as StackOverflow).

The model shown in Figure 1 has two dimensions: expertise and outlet control. Both dimensions run between extremes "unknown" and "known". Expertise is the extent to which the authority and knowledge of the producer of the content can be determined. Outlet control is the extent to which content is produced, moderated or edited in conformance with explicit and transparent knowledge creation criteria. Rather than having discrete bands, the gradation in both dimensions is on a continuous range between known and unknown, producing the shades of GL.

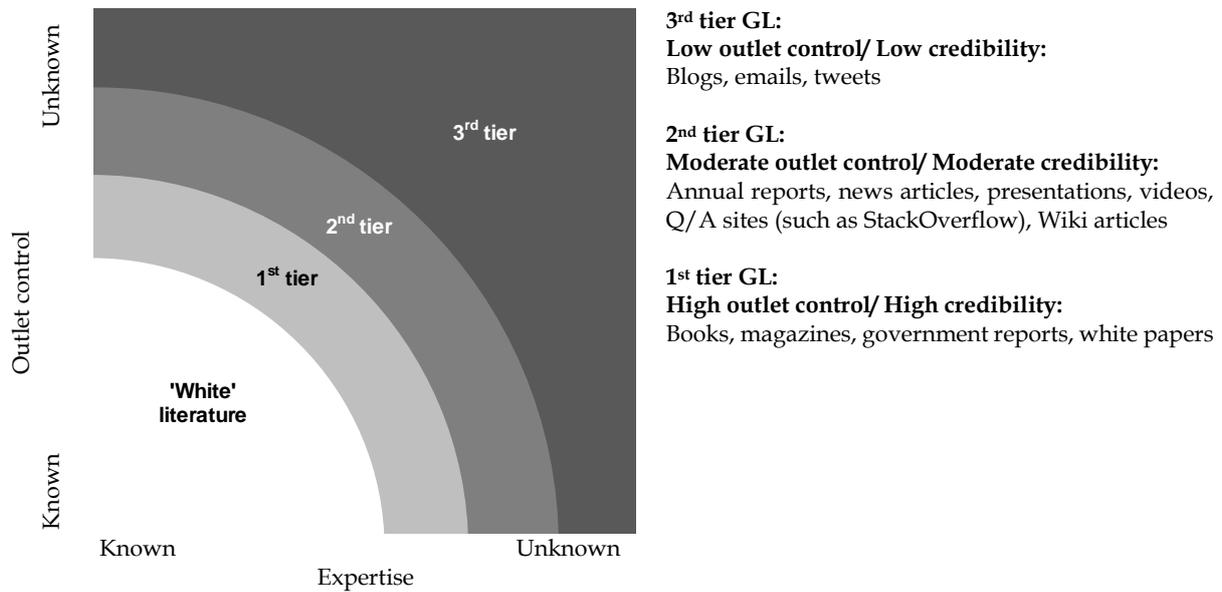

**Figure 1- "Shades" of grey literatures (from [17])**

The "shades" of grey model shown in Figure 1 is quite consistent with Table 1 showing the spectrum of the 'white', 'grey' and 'black' literature from another source [25]. The 'white' literature is visible in both Figure 1 and Table 1 and the means the source where both expertise and outlet control are fully known. 'Grey' literature according to Table 1 corresponds mainly to the 2nd tier in Figure 1 with moderate outlet control and credibility. For SE, we add Q/A sites like StackOverflow to the 2nd tier. 'Black' literature finally corresponds to ideas, concepts and thoughts. As blogs, but also emails and tweets mainly refer to ideas, concepts or thoughts they are in the 3rd tier. However, there are even "shades" of grey in the classification and depending on the concrete content a specific type of grey literature can be in a different tier than shown in Figure 1. For instance, if a presentation (or a video, which is often linked to a presentation) is about new ideas, then it would fall into the 3rd tier.

**Table 1- Spectrum of the 'white', 'grey' and 'black' literature (from [25])**

| 'White' literature | 'Grey' literature | 'Black' literature |
|---|---|---|
| Published journal papers | Preprints | Ideas |
| Conference proceedings | e-Prints | Concepts |
| Books | Technical reports | Thoughts |
| | Lectures | |
| | Data sets | |
| | Audio-Video (AV) media | |
| | Blogs | |

Due to the limited control of expertise and outlet in GL, it is important to also identify GL producers. According to [25] following GL producers were identified: (1) Government departments and agencies (i.e., in municipal, provincial, or national levels); (2) Non-profit economic and trade organizations; (3) Academic and research institutions; (4) Societies and political parties; (5) Libraries, museums, and archives; (6) Businesses and corporations; and (7) Freelance individuals, i.e., bloggers, consultants, and web 2.0 enthusiasts. For SE, it might in addition also be relevant to distinguish different types of companies, e.g. startups versus established organizations, or different governmental organizations, e.g. military versus municipalities, producing GL. From a highly-cited paper from the medical domain [26], we can see that GL searches can go far beyond simple Google searches as the authors searched "*44 online resource and database websites, 14 surveillance system websites, nine regional harm reduction websites, three prison literature databases, and 33 country-specific drug control agencies and ministry of health websites*". That paper highlighted the benefits of the GL by pointing out that 75% to 85% of their results were based on data sourced from the GL.

## 2.2 Different types of secondary studies

A secondary study is a study of studies. A secondary study does usually not generate any new data from a "direct" (primary) research study, instead it analyses a set of primary studies and usually seeks to aggregate the results from these in order to provide stronger forms of evidence about a particular phenomenon [27]. In the research community, a secondary study is sometimes also called a "survey paper" or a "review paper" [28, 29]. There are different types of secondary studies. For example, a review of 101 secondary studies in software testing [29] classified secondary studies into the following types: regular surveys, systematic literature reviews (SLR), systematic literature mappings (SLM or SM).

The number of secondary studies in many research fields has grown very rapidly in recent years. To get a sense for the popularity of systematic reviews, we searched for the term "systematic review" in paper titles in the Scopus search engine. As of this writing (April 24, 2018), this phrase returned 86,525 papers. We also did the same search, but wanted to focus only on the SE discipline. To do so in an automated manner, we specified in the search criteria the term "software" appears in "source title", i.e., venue (journal or conference) name. This approach was used in several recent bibliometric studies, e.g., [30-32], and was shown to be a precise way to automatically search for SE papers in Scopus. The search for "systematic review" in SE paper titles returned 401 papers as of this writing (April 2018).

In general, secondary studies are of high value both for SE practice and research. For instance, when asked about the benefit of a recent survey paper on testing embedded software, a practitioner tester mentioned that [33]: "*There are a lot of studies in the pool of this review study, which would benefit us in choosing the best methods to test embedded software systems. I think review studies such as this one could be very beneficial for companies like ours*". Furthermore, a recent tertiary study on software testing (a SLR of 101 secondary studies in software testing) [29] stresses the important role of secondary studies in SE in general and software testing in particular. It compared citations of secondary with citations of primary studies. The study found that, citation metrics to the secondary studies were higher than the papers in the pool of three SM studies (web testing [34], GUI testing [35] and UML-SPE [36]). This suggests that the research community has already recognized the value of secondary studies, as secondary studies are cited on average higher than regular primary studies. Thus, it appears that if a secondary study (or a MLR) is conducted with interesting and "useful" RQs, it could bring value and benefit to practitioners and researchers.

As publishing various types of GL besides formal scientific literature is becoming more popular and widespread, adapted types of secondary studies, e.g., Multivocal Literature Reviews (MLR), are becoming popular as well. Therefore, respective guidelines for Multivocal Literature Reviews, that take GL into account, are needed. This article provides guidelines to perform newer types of secondary studies to ensure effective/efficient execution of such studies and high quality of reported reviews.

To better characterize secondary studies in SE, we categorize the types of systematic secondary studies in SE and briefly discuss their similarities, difference and relationships. Based on the review of the literature and our studies in this area, e.g., [29], we categorize secondary studies in SE into six types, i.e., Systematic Literature Mappings (SLM), Systematic Literature Review (SLR), Grey Literature Mapping (GLM), Grey Literature Review (GLR), Multivocal Literature Mapping (MLM), and Multivocal Literature Review (MLR) (see Figure 2).

As we specify in Figure 2, the differentiation factors of six types of systematic secondary studies are: types of analysis, and types of sources under study. For example, the difference between an MLR and an SLR is the fact that, while SLRs use as input only academic peer-reviewed articles, MLRs in addition also use sources from the GL, e.g., blogs, white papers, videos and web-pages [18].

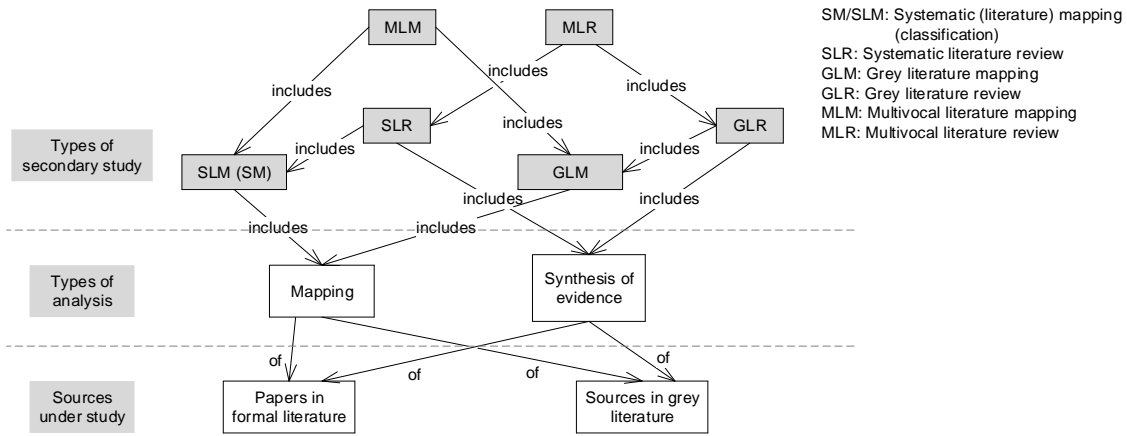

**Figure 2-Relationship among different types of systematic secondary studies**

Another type of literature reviews are GLR. As the name implies, they only consider GL sources in their pool of reviewed sources. Many GLR studies have also appeared in other disciplines, e.g., in medicine or social science [37-40]. For example, a GLR of special events for promoting cancer screenings was reported in [37]. To better understand and characterize the relationship between SLM, GLM and MLR studies, we visualize their relationship as a Venn diagram in Figure 3. The same relationship holds among SLR, GLR and MLM studies (see Figure 2). As Figure 3 clearly shows, an MLR in a given subject field is a union of the sources that would be studied in an SLR and in a GLR of that field. As a result, an MLR, in principle, is expected to provide a more complete picture of the evidence as well as the state-of-the-art and -practice in a given field than an SLR or a GLR (we will discuss this aspect more in the next sub-section by rephrasing some results of our previous work in [41]).

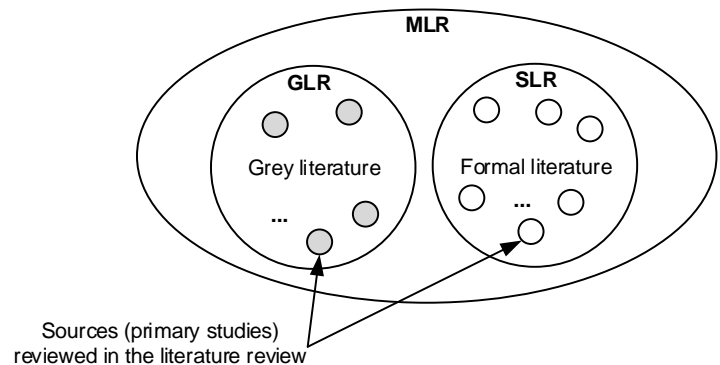

**Figure 3- Venn diagram showing the relationship of SLR, GLR and MLR studies**

Studies from all six types shown in Figure 2 have started to be appear in SE, e.g., a recent GLR paper [42] was published on the subject of choosing the right test automation tools. A Multivocal Literature Mapping (MLM) is conducted to classify the body of knowledge in a specific area, e.g., a MLM on software test maturity assessment and test process improvement [43]. Similar to the relationship of SLM and SLR studies [22], a MLM can be extended by follow-up studies to a Multivocal Literature Review (MLR) where an additional in-depth analysis or qualitative coding of the issues and evidence in a given subject is performed, e.g., [44].

## 2.3 Benefits of and need for including grey literature in review studies (conducting MLRs)

Our previous work [41] explored the need for MLRs in SE. Our key findings indicated that (1) GL can give substantial benefits in certain areas of SE, and that (2) the inclusion of GL brings forward certain challenges as evidence in them is often experience and opinion based. We found examples that numerous practitioner sources had been ignored in previous SLRs and we think that missing such information could have profound impact on steering research directions. On the other hand, in that paper, we demonstrated the information gained when making an MLR. For example, the MLR on the subject of deciding when and what to automate in testing [44] would have missed a lot of expertise from test engineers if we had not included the GL, see Figure 4.

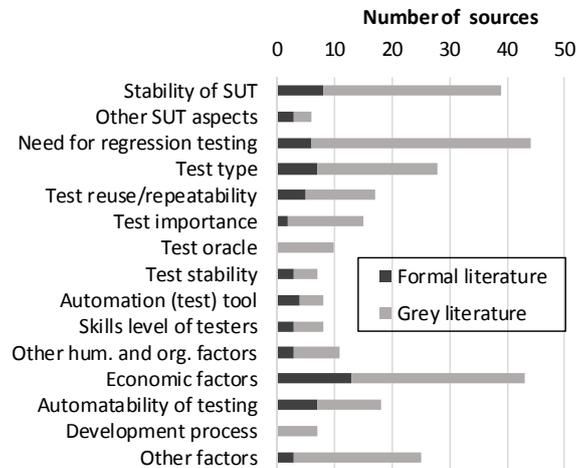

**Figure 4-An output of the MLR on deciding when and what to automate in testing (MLR-AutoTest)**

Also, in other domains (e.g., in educational sciences), a key benefit of MLRs has been "*closing the gap between academic research and professional practice*" [45], which was reported as early as in 1991. We have also observed in the execution and usage of review results in a few MLRs that we have been involved in, e.g., [43, 44]. One main reason to conduct both those MLR studies [43, 44] were the real-world needs in industrial settings that we had w.r.t. the topics of these two MLRs: When and what to automate in software testing in the case of [44], and software test maturity and test process improvement in the case of [43]. As reported in [46-49], we were faced with the challenge of systematically decision when (in the lifecycle) and what (which test cases) to automate in several industrial contexts. The MLR study that we conducted [44] synthesized both the state-of-the art and the state-of−practice to ensure that we would benefit from both research and also industrial knowledge to answer the challenging questions (see Figure 4). In a recent study [46], we used the results of the MLR [44] in practice and found the results very useful. We had a similar positive experience in using results from the other MLR [43] in our recent projects in software test maturity and test process improvement.

It should be highlighted that we are not advocating that all SLRs in SE should include GL and become MLRs. But instead, as we explain in Section 4.2, researchers considering to conduct an SLR from formal literature only in a given SE topic, should assess whether "broadening" the scope and including GL would add value and benefits to the review study and, only when the answer to those questions is positive, they should plan an MLR instead of an SLR. We will review the existing guidelines for those decisions in Section 4.2 and will adopt them to the SE context. Finally, it should be noted that including the GL in review studies is not always straightforward or advantageous [50]. There are some drawbacks as well, e.g., lower quality reporting on particularly when describing research methodology. Thus, careful considerations should be taken in different steps of an MLR study to be aware of such drawbacks (details in Section 4-6).

## 2.4 GL and MLRs in SE

While extensive GL is available in the field of SE and the volume of GL in SE is clearly expanding on a very rapid pace (e.g., in blogs and free online books), little effort has been made to utilize such a knowledge in SE research. Recently small steps in this direction have been made by Rainer who reported in [51] a preliminary framework and methodology based on "argumentation" theory [52] to identify, extract and structure SE practitioners' evidence, inference and beliefs. The authors argued that practitioners use (factual) stories, analogies, examples and popular opinion as evidence, and use that evidence in defeasible reasoning to justify their beliefs in GL sources (such as blogs) and to rebut the beliefs of other practitioners. Their paper [51] showed that the presented framework, methodology and examples could provide a foundation for SE researchers to develop a more sophisticated understanding of, and appreciation for, practitioners' defeasible evidence, inference and belief. We will utilize some inputs from the study of Rainer [51] in development of our guidelines, especially for data synthesis (Section 5.5).

MLRs have recently started to emerge as a type of secondary study in SE. The "multivocal" terminology has recently started to appear in SE. Based on a literature search, we found several MLR studies in SE [18, 43, 44, 53-59]. We list those MLRs in Table 2 together with their topics, years of publication and the information about the number of sources from the formal literature and the GL as well as the ratio (%) of GL in the pool.

From Table 2, one can see that MLRs are a recent trend in SE, as more researchers are seeing the benefit in conducting them (as discussed above). About nine MLRs have been published in SE between 2015 and 2018. As Table 2 shows, scale of the listed MLRs vary w.r.t. the number of sources reviewed. While [58] studied serious games for software process standards education on a small set of 7 sources (of which only 1 was from the GL), [55] reviewed relationship of DevOps to agile, lean and continuous deployment on a large set of 234 sources (of which 201 were from the GL). Ratio of GL in the pools of the MLRs also vary, from 14.3% in [58] to 85.9% in [55], which of course is due to the nature of the topic under study, i.e., relationship of DevOps to agile, lean and continuous deployment seems to be a topic very active in the industry compared to academia.

In some software engineering MLR's, an SLR has been performed prior to undertaking the grey literature review of the MLR or the authors' prior work has had an existing SLR, e.g., [18, 53] [55, 58] (see Table 2). However, there are papers that have done parallel SLR and grey literature reviews, e.g., [43, 44, 54, 56, 59]. Some have also combined MLRs with interviews, e.g., [18, 55]. There are also some papers that have only done grey literature review, e.g., [42] [60]. It is hard to reason on the order as it depends on the goal and the existing body of academic and practitioner work.

**Table 2- List of MLRs in SE (sorted by year of publication)**

| Year | Topic and Reference | Total - % of GL in the pool | Literature used for MLR methodology and a brief summary of MLR process. |
|---|---|---|---|
| 2013 | An exploration of technical debt [18] | 35 - 100% | This paper used MLR information from [12]. They used previously performed SLR for designing a grey literature review. After grey literature, also interviews were done to collect primary data. They used top hits 50 from Google and performed two iterations of searches were the second iteration included new terms found in the first iteration. Quality filtering was done case-by-case. |
| 2015 | iOS applications testing [53] | 21- 42.9% | This paper used MLR information from [12]. This paper first performed academic searches (SLR). Then it used keywords from academic search that were modified for the grey literature search. The paper studied the first 50 hits provided by Google search engine. Topic and quality based filtering was done for the MLR. |
| 2016 | When and what to automate in software testing [44] | 78 - 66.7% | This (MLR-AutoTest) is one of our prior works and it references multiple prior works about MLR and including grey literature yet the depth does not match this paper as that was not a methodological paper. This paper is used as an example throughout this paper. |
| 2016 | Gamification of software testing [54] | 20 - 70.0% | This is one of our prior works that uses the same strategy as in [44] but in general the approach is more limited as it was only a short paper for a conference rather than journal paper. |
| 2016 | Relationship of DevOps to agile, lean and continuous deployment [55] | 234 - 85.9% | This paper used MLR information from [12], [18], and [44] to device a search strategy. The paper combined three data source as it performed it first performed grey literature review, then did an update of an SLR and finally collected primary information from practitioners. The paper makes no mention how SLR and grey literature search are linked. First 230 hits of Google search engine were included as it was determined that hits below that were mostly job adds. Topic and quality based filtering was done. |
| 2016 | Characterizing DevOps [56] | 43 - 44.2% | This paper used MLR information from [12]. They searched Google (grey literature) and Google Scholar (MLR) no indication is given whether one was searched before the other. Data collection and extraction was interleaved and search was stopped when no additional data could be extracted from new sources. |
| 2017 | Threat intelligence sharing platforms of software vendors [57] | 22 - NA | This paper used MLR information from [12, 18], and our previous work [41]. The paper used 9 academic search engines and 2 search engines. No details on stopping criteria were given. Quality criteria was used for filtering. |
| 2017 | Serious games for software process standards education [58] | 7 - 14.3% | In this paper, scientific searches were done first. Only using the scientific search results grey literature search was performed. It consisted of two steps both using the academic primary studies: 1) for backward and forward snowballing, and 2) for studying the publication list of each academic author to find all the works the authors have performed in this area. |

| 2017 | Software test maturity and test process improvement [43] | 181 - 28.2% | This is one of our prior works that uses the same strategy as in [44]. |
|------|------|------|------|
| 2018 | Smells in software test code [59] | 166 - 27.7% | This is one of our prior works that uses the same strategy as in [44]. |

Other SLRs have also included the GL in their reviews and have not used the "multivocal" terminology, e.g., [61]. A 2012 MSc thesis [50] explored the state of including the GL in the SE SLRs and found that the ratio of grey evidence in the SE SLRs was only about 9%, and the GL evidence concentrated mostly in the recent past (~48% between the years 2007-2012). Furthermore, using GL as data has been described as a case study, as was done in a 2017 paper investigating pivoting in software start-up companies [60].

## 2.5 Lack of existing guidelines for conducting MLR studies in SE

Although, the existing SLR guidelines (e.g., those by Kitchenham and Charters [22]) have briefly discussed the idea of including GL sources in SLR studies, most SLRs, published so far in SE, have not actually included GL in their studies. A search for the word "grey" in the SLR guideline by Kitchenham and Charters [22] just returns two hits, which we cite below:

"*Other sources of evidence must also be searched (sometimes manually) including:*

- *Reference lists from relevant primary studies and review articles*
- *Journals (including company journals such as the IBM Journal of Research and Development), grey literature (i.e. technical reports, work in progress) and conference proceedings*
- *Research registers*
- *The Internet*"

And:

"*Many of the standard search strategies identified above are used, …, including:*

- *Scanning the grey literature*
- *Scanning conference proceedings*"

While guidelines for SLR studies, e.g., [22], and SM studies [3, 21], could be useful for conducting MLRs, they do not provide specific guidance on how to treat GL in particular, since GL sources should be assessed differently in some steps compared to formal literature, e.g., quality assessment (as we discuss in Section 5.3).

Table 2 present analysis which shows that first works in SE have mainly cited [12] from education sciences when presenting their MLR process. More recent works have cited already existing MLR studies in SE such as [18] and [44] when presenting the MLR process. In the papers of Table 2, the treatment of MLR methodology is quite brief typically, 2-4 paragraphs, as they are not methodological papers. Our guidelines offer much broader coverage of MLR literature than any of the previous MLR studies in SE.

To summarize a lack of MLR guidelines in the SE literature can be stated. In particular, two papers explicitly discussed this shortage as follows: "*there are no systematic guidelines for conducting MLRs in computer science*" [57] and "*There is no explicit guideline for collecting ML [multivocal literature]*" [55]. We are addressing that need in this paper.

## 3 AN OVERVIEW OF THE GUIDELINES AND ITS DEVELOPMENT

In Section 3.1, we explain how we developed the guidelines and Section 3.4 provides an overview of the guidelines.

### 3.1 Developing the guidelines

In this section, we discuss our approach to deriving the guidelines for including the GL and conducting MLRs in SE. Figure 5 shows an overview of our methodology. Four sources are used as input in the development of MLR guidelines:

(1) A survey of 24 MLR guidelines and experience papers in other fields;
(2) Existing guidelines for SLR and SM studies in SE, notably the popular SLR guidelines by Kitchenham and Charters [22];

(3) The experience of the authors in conducting several MLRs [43, 44, 54, 62] and one GLR [42]; and

(4) A recent study by Rainer [51] on using argumentation theory to analyze software practitioners' defeasible evidence, inference and belief

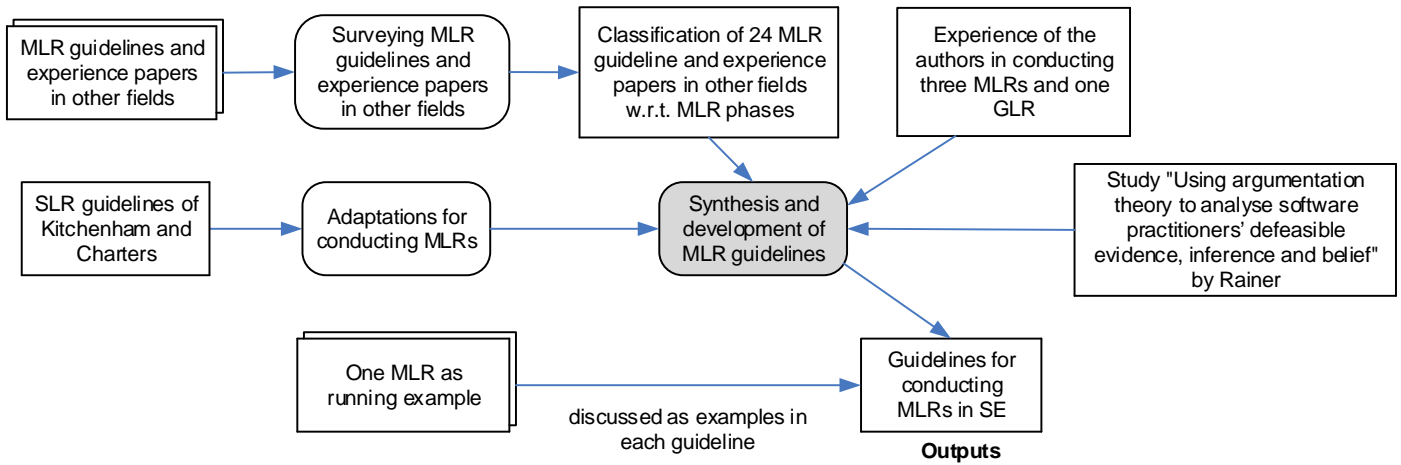

**Figure 5-An overview of our methodology for developing the guidelines in this paper**

There are several guidelines for SLR and SM studies in SE available [3, 21, 22, 63]. Yet, they mostly ignore the utilization of GL, as discussed in Section 2.4. Therefore, we see that our guidelines fill a gap by raising the importance of including GL in review studies in SE and by providing concrete guidelines with examples on how to address and include GL in review studies.

As shown in Figure 5, we also used our own expertise from our recently-published MLRs [43, 44, 54, 62] and one GLR [42]. Additionally, our experience includes several SLR studies, e.g., [34-36, 64-68].

## 3.2 Surveying MLR guidelines in other fields

As shown in Figure 5, one of the important sources used as input in the development of our MLR guidelines was a survey of MLR guidelines and experience papers in other fields. Via a systematic survey, we identified 24 such papers and conducted a review of those studies. The references of those 24 papers are as follows: [12-15, 17, 19, 20, 25, 45, 50, 69-82].

Each of those 24 MLR guideline and experience papers provided guidelines for one or several phases of a MLR: (1) decision to include GL in review studies, (2) MLR planning, (3) search process, (4) source selection (inclusion/exclusion), (5) source quality assessment, (6) data extraction, (7) data synthesis, (8) reporting the review (dissemination), and (9) any other type of guideline. In the rest of this paper, we have synthesized those guidelines and have adopted them to the context of MLRs in SE by consolidating them with our own experience in MLRs.

Figure 6 shows the number of papers from the set of those 24 papers, per each phase of a MLR. For example, 14 of those 24 papers provided guidelines for the search process of conducting a MLR. Details about this classification of MLR guideline papers can be found in an online source [83] available at goo.gl/b2u1E5.

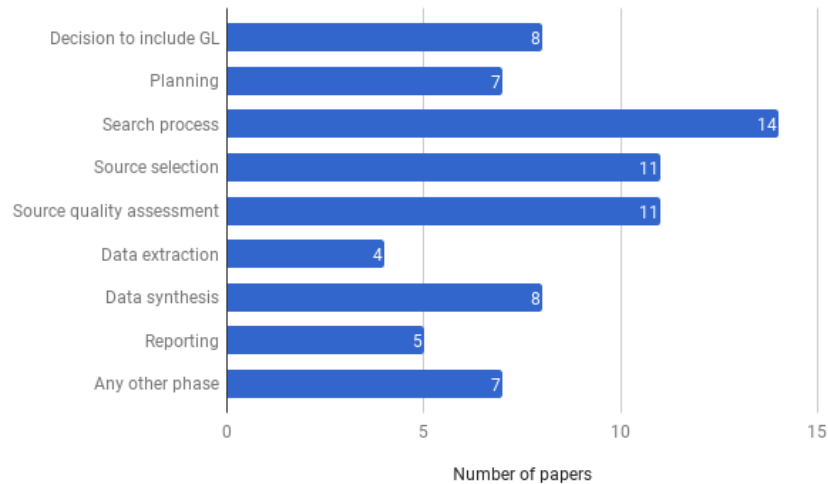

**Figure 6-Number of papers in other fields presenting guidelines of different activities of MLRs (details can be found in [83])**

### 3.3 Running example

We selected one MLR [44], on deciding when and what to automate in testing, as the running example and, we refer to it as MLR-AutoTest in the remainder of this guideline paper. When we present guidelines for each step of the MLR process in the next sections, we discuss whether and how the respective step and the guidelines were implemented in MLR-AutoTest.

Since we developed the guidelines presented in this paper after conducting several MLR studies, and based on our accumulated experience, it could be that certain steps of the guideline were not systematically applied in MLR-AutoTest. In such cases, we will discuss how the guidelines of a specific step "should have been" conducted in that MLR. After all, working with GL has been a learning experience for all of the three authors.

### 3.4 Overview of the guidelines

From the SLR guidelines of Kitchenham and Charters [22], we adopt three phases (1) planning the review, (2) conducting the review, and (3) reporting the review for conducing MLRs, since we have found them to be well classified and applicable to MLRs. The corresponding phases of our guidelines are presented in Sections 3, 4 and 5, respectively. There are also sub-steps for each phase as shown in Table 3. To prevent duplication, we do not repeat all steps of the SLR guidelines [22] when they are the same for conducting MLRs, but only present the steps that are different for conducting MLRs. Therefore, our guidelines focus mainly on GL sources as handling sources from the formal literature is already covered by the SLR existing guidelines. Integrating both types of sources in an MLR is usually straightforward, as per our experience in conducting MLRs [43, 44, 54, 62].

**Table 3- Phases of the Kitchenham and Charters' SLR guidelines (taken from page 6 of [22])**

| Phase | Steps |
|---|---|
| Planning the review | • Identification of the need for a review<br>• Commissioning a review<br>• Specifying the research question(s)<br>• Developing a review protocol<br>• Evaluating the review protocol |
| Conducting the review | • Identification of research<br>• Selection of primary studies<br>• Study quality assessment<br>• Data extraction and monitoring<br>• Data synthesis |
| Reporting the review | • Specifying dissemination mechanisms |

| | • Formatting the main report<br>• Evaluating the report |
|---|---|

## 4 PLANNING A MLR

As shown in Figure 7, the MLR planning phase consists of the following two phases: (1) Establishing the need for an MLR in a given topic, and (2) Defining the MLR's goal and raising its research questions (RQs). In this section, these two steps are discussed.

### 4.1 A typical process for MLR studies

We illustrate a typical MLR process in Figure 7. As one can see, this process is based on the SLR process as presented in Kitchenham and Charters' guidelines [22] and has been adapted to the context of multivocal literature reviews. Our figure visualizes the process, for better understandability, and we have extended it to make it suitable for MLRs. In Figure 7, we have also added the numbers of the sections, where we cover guidelines for specific process steps, to ease traceability between this process and the paper text. The process can also be applied to structure a protocol on how the review will be conducted. An alternative way to develop a protocol for MLRs is to apply the standard structure of a protocol for SLRs [27] and to consider the guidelines provided in this paper as specific variation points on how to consider GL. We believe that having a baseline process (template) from which other researchers can make their extensions/revisions could provide a semi-homogenous process for conducting MLRs, and thus provide the first of our set of guidelines as follows:

> ⇒ **Guideline 1:** The provided typical process of an MLR can be applied to structure a protocol on how the review will be conducted. Alternatively, the standard protocol structure of SLR in SE can be applied and the provided guidelines can be considered as variation points.

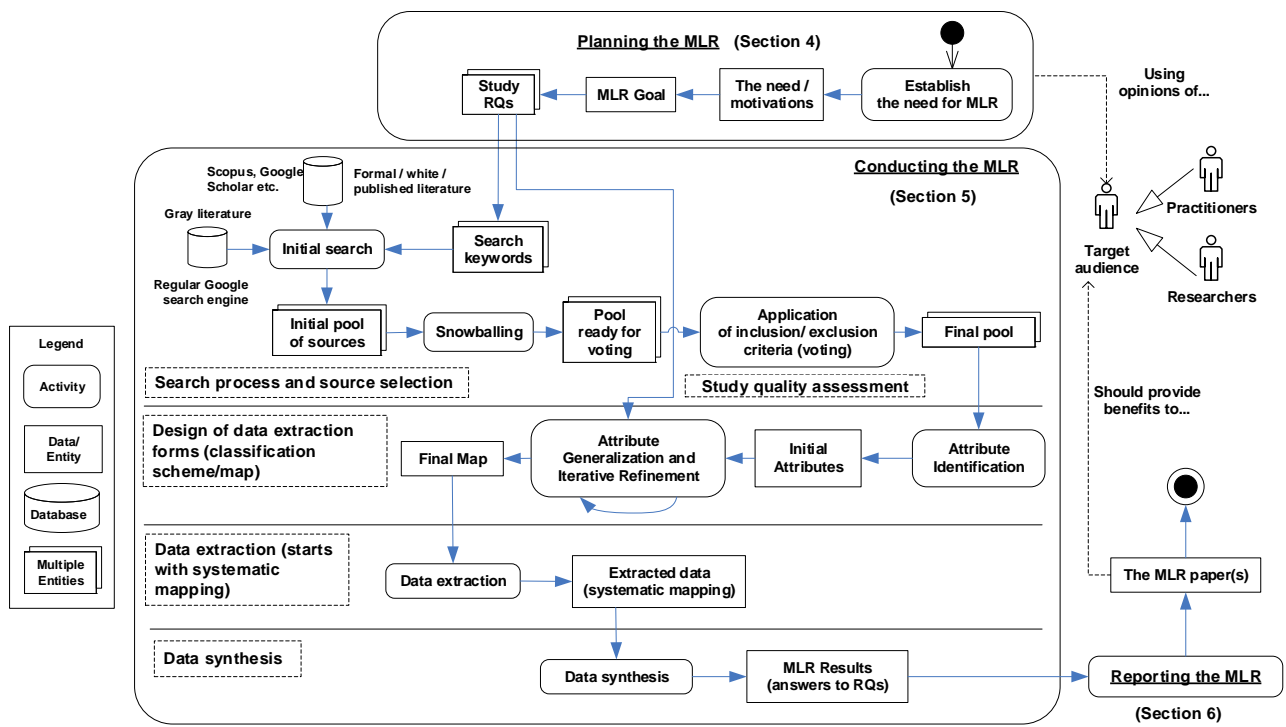

**Figure 7-An overview of a typical MLR process**

### 4.2 Raising (motivating) the need for a MLR

Prior to undertaking an SLR or an MLR, researchers should ensure that conducting a systematic review is necessary. In particular, researchers should identify and review any existing reviews of the phenomenon of interest [22]. We also think that conductors of an MLR or SLR should pay close attention to ensure the usefulness of an MLR for its intended audience, i.e., researchers and/or practitioners, as early as its planning phase for defining of its scope, goal and review questions [3].

For example, the motivation of the MLR-AutoTest completely started from our industry-academia collaboration on test automation. Our industry partners had challenges to systematically decide when and what to automate in testing, e.g., [46-49], and thus we felt the real industrial need to conduct the MLR-AutoTest. Furthermore, since we found many GL on that topic, conducting a MLR was seen much more logical than a SLR of academic sources. This brings us to an important guideline about motivating the need for a MLR:

> **Guideline 2:** Identify any existing reviews and plan/execute the MLR to explicitly provide usefulness for its intended audience (researchers and/or practitioners).

While establishing the need for a review, one should assess whether to perform SLR, GLR or MLR or their mapping study counterparts, see Figure 2. Note that the question of whether or not to include the GL is the same as whether or not to conduct an MLR instead of an SLR. If the answer to that question is negative, then the next question is whether or not to conduct an SLR instead, which has been covered by their respective guidelines [3, 22]. Several MLR guidelines from other fields have addressed the decision whether to include the GL and conduct an MLR instead of an SLR. For example, they provide the following suggestions.

- GL provides "*current*" perspectives and complements gaps of the formal literature [25].
- Including GL may help avoiding publication bias. Yet, the GL that can be located may be an unrepresentative sample of all unpublished studies. [19]
- Decision to include GL in an MLR was a result of consultation with stakeholders, practicing ergonomists, and health and safety professionals [80].
- If GL were not included, the researchers thought that an important perspective on the topic would have been lost [80]., and we observed a similar situation in the MLR-AutoTest, see Figure 4.

Importantly, we found two checklists whether to include GL in an MLR. A checklist from [81] includes six criteria. We want to highlight that according to [81] GL is important when context has a large effect on the implementation and the outcome which is typically the case in SE [84, 85]. We think that GL may help in revealing how SE outcomes are influenced by context factors like the domain, people, or applied technology. Another guideline paper [17] suggests including GL in reviews when relevant knowledge is not reported adequately in academic articles, for validating scientific outcomes with practical experience, and for challenging assumptions in practice using academic research. This guideline also suggests excluding GL from the reviews of relatively mature and bounded academic topics. In SE, this would mean topics such as the mathematical aspects of formal methods which are relatively bounded in the academic domain only, i.e., one would not find too many practitioner-generated GL on this subject.

Based on [81] and [17] and our experience, we present our synthesized decision aid in Table 4. Note that, one or more "yes" responses suggest the inclusion of GL. Items 1 to 5 are adopted from prior sources [17, 81], while items 6 and 7 are added based on our own experience in conducting MLRs. For example, item #3 originally [17, 81] was: "*Is the context important to the outcome or to implementing intervention?*". We have adopted it as shown in Table 4. It is increasingly discussed in the SE community that "contextual" information (e.g., what approach works for whom, where, when, and why?) [86-89] are critical for most of SE research topics and shall be carefully considered. Since GL sometimes provide contextual information, including them and conducting a MLR would be important. It is true that question 3 would (almost) always be yes for most SE topics, but we still would like to keep it in the list of questions, in case.

In Table 4, we also apply the checklist of [81] to our running MLR example (MLR-AutoTest) as an "a-posteriori" analysis. While some of the seven criteria in this list may seem subjective, we think that a team of researchers can assess each aspect objectively. For MLR-AutoTest, the sum of "Yes" answers is seven as all items have "Yes" answers. The larger the sum the higher is the need for conducting an MLR on that topic.

**Table 4-Questions to decide whether to include the GL in software engineering reviews**

| # | Question | Possible answers | MLR-AutoTest |
|---|---|---|---|
| 1 | Is the subject "complex" and not solvable by considering only the formal literature? | Yes/No | Yes |
| 2 | Is there a lack of volume or quality of evidence, or a lack of consensus of outcome measurement in the formal literature? | Yes/No | Yes |
| 3 | Is the contextual information important to the subject under study? | Yes/No | Yes |

| 4 | Is it the goal to validate or corroborate scientific outcomes with practical experiences? | Yes/No | Yes |
| 5 | Is it the goal to challenge assumptions or falsify results from practice using academic research or vice versa? | Yes/No | Yes |
| 6 | Would a synthesis of insights and evidence from the industrial and academic community be useful to one or even both communities? | Yes/No | Yes |
| 7 | Is there a large volume of practitioner sources indicating high practitioner interest in a topic? | Yes/No | Yes |

*Note: One or more "yes" responses suggest inclusion of GL.*

⇒ **Guideline 3:** The decision whether to include the GL in a review study and to conduct an MLR study (instead of a conventional SLR) should be made systematically using a well-defined set of criteria/questions (e.g., using the criteria in Table 4).

### 4.3 Setting the goal and raising the research questions

The SLR guidelines of Kitchenham and Charters [22] state that specifying the RQs is the most important part of any systematic review. To make the connection among the review's goal, research (review) questions (RQs) as well as the metrics to collect in a more structured and traceable way, we have often made use of the Goal-Question-Metric (GQM) methodology [90] in our previous SM, SLR and MLR studies [34-36, 64-68]. In fact, the RQs drive the entire review by affecting the following aspects directly:

- The search process must identify primary studies that address the RQs
- The data extraction process must extract the data items needed to answer the RQs
- The data analysis (synthesis) phase must synthesize the data in such a way that the RQs are properly answered

Table 5 shows the RQs raised in the example MLR. MLR-AutoTest raised four RQs and several sub-RQs under some of the top-level RQs. This style was also applied in many other SM and SLR studies to group the RQs in categories.

**Table 5- The RQs raised in the example MLR (MLR-AutoTest)**

| MLR study | RQs |
|---|---|
| MLR-AutoTest | • RQ 1-Mapping of sources by contribution and research-method types:<br>    ○ RQ 1.1- How many studies present methods, techniques, tools, models, metrics, or processes for the when/what to automate questions?<br>    ○ RQ 1.2- What type of research methods have been used in the studies in this area?<br>• RQ 2-What factors are considered in the when/what questions?<br>• RQ 3- What tools have been proposed to support the when/what questions?<br>• RQ 4- What are attributes of those systems and projects?<br>    ○ RQ 4.1- How many software systems or projects under analysis have been used in each source?<br>    ○ RQ 4.2- What are the domains of the software systems or projects under analysis that have been studied in the sources (e.g., embedded, safety-critical, and control software)?<br>    ○ RQ 4.3- What types of measurements, in the context of the software systems under analysis, to support the when/what questions have been provided? |

RQs should also match specific needs of the target audience. For example, in the planning phase of the MLR-AutoTest, we paid close attention to ensure the usefulness of that MLR for its intended audience (practitioners) by raising RQs which would benefit them, e.g., what factors should be considered for the when/what questions?

Another important criteria in raising RQs is to ensure that they are as objective and measurable as possible. Open-ended and exploratory RQs are okay but RQs should not be fuzzy or vague.

⇒ **Guideline 4:** Based on your research goal and target audience, define the research (or "review") questions (RQs) in a way to (1) clearly relate to and systematically address the review goal, (2) match specific needs of the target audience, and (3) be as objective and measurable as possible.

Based on our own experience, it would also be beneficial to be explicit about the proper type of the raised RQs. Easterbrook et al. [91] provide a classification of RQ types that we used to classify a total of 267 RQs studied in a pool of 101 literature reviews in software testing [29]. The adopted RQ classification scheme [91] and examples RQs from the reviewed studies in [29] are shown in Table 6. The findings of the study [29] showed that, in its pool of studies, descriptive-classification RQs were the most popular by large margin. The study [29] further reported that there is a shortage or lack of RQs in types towards the bottom of the classification scheme. For example, among all the studies, no single RQ of type Causality-Comparative Interaction or Design was raised.

For MLR-AutoTest, as shown in Table 5, all of its four RQs were of type "descriptive-classification". If the researchers are planning an MLR with the goal of finding out about "relationships", "causality", or "design" or certain phenomena, then they should raise the corresponding type of RQs. We would like to express the need for RQs of types "relationships", "causality", or "design" in future MLR studies in SE. However, we are aware that the primary studies may not allow such questions to be answered.

⇒ **Guideline 5:** Try adopting various RQ types (e.g., those in Table 6) but be aware that primary studies may not allow all question types to be answered.

**Table 6- A classification scheme for RQs as proposed by [91] and examples RQs from a tertiary study [29]**

| RQ category | Sub-category | Example RQs |
|---|---|---|
| Exploratory | Existence | Does X exist?<br>• Do the approaches in the area of product lines testing define any measures to evaluate the testing activities? [S2]<br>• Is there any evidence regarding the scalability of the meta-heuristic in the area of search-based test-case generation?<br>• Can we identify and list currently available testing tools that can provide automation support during the unit-testing phase? |
| | Description-Classification | What is X like?<br>• Which testing levels are supported by existing software-product-lines testing tools?<br>• What are the published model-based testing approaches?<br>• What are existing approaches that combine static and dynamic quality assurance techniques and how can they be classified? |
| | Descriptive-Comparative | How does X differ from Y?<br>• Are there significant differences between regression test selection techniques that can be established using empirical evidence? |
| Base-rate | Frequency Distribution | How often does X occur?<br>• How many manual versus automated testing approaches have been proposed?<br>• In which sources and in which years were approaches regarding the combination of static and dynamic quality assurance techniques published?<br>• What are the most referenced studies (in the area of formal testing approaches for web services)? |
| | Descriptive-Process | How does X normally work?<br>• How are software-product-lines testing tools evolving?<br>• How do the software-product lines testing approaches deal with tests of non-functional requirements?<br>• When are the tests of service-oriented architectures performed? |
| Relationship | Relationship | Are X and Y related?<br>• Is it possible to prove the independence of various regression-test-prioritization techniques from their implementation languages? |
| Causality | Causality | Does X cause (or prevent) Y?<br>• How well is the random variation inherent in search-based software testing, accounted for in the design of empirical studies?<br>• How effective are static analysis tools in detecting Java multi-threaded bugs and bug patterns?<br>• What evidence is there to confirm that the objectives and activities of the software testing process defined in DO-178B provide high quality standards in critical embedded systems? |
| | Causality-Comparative | Does X cause more Y than does Z?<br>• Can a given regression-test selection technique be shown to be superior to another technique, based on empirical evidence?<br>• Are commercial static-analysis tools better than open-source static-analysis tools in detecting Java multi-threaded defects?<br>• Have different web-application-testing techniques been empirically compared with each other? |
| | Causality-Comparative Interaction | Does X or Z cause more Y under one condition but not others?<br>• There were no such RQs in the pool of the tertiary study [29] |
| Design | Design | What's an effective way to achieve X? |

| | | • There were no such RQs in the pool of the tertiary study [29] |
|---|---|---|

## 5 CONDUCTING THE REVIEW

Once an MLR is planned, it shall be conducted. This section is structured according to five phases of conducting an MLR:

- Search process (Section 5.1)
- Source selection (Section 5.2)
- Study quality assessment (Section 5.3)
- Data extraction (Section 5.4)
- Data synthesis (Section 5.5)

### 5.1 Search process

Searching either formal or GL is typically done via means of using defined search strings. Defining the search strings is an iterative search process, where the initial exploratory searches reveal more relevant search strings. Literature can also be searched via a technique called "snowballing" [92], where one follows citations either backward or forward from a set of seed papers. Here we highlight the differences between searching in formal literature versus GL.

### 5.1.1 Where to search

Formally-published literature is searched via either broad-coverage abstract databases, e.g., Scopus, Web of Science, Google Scholar or from full-text databases with more limited coverage, e.g., IEEE Xplore, ACM digital library, or ScienceDirect. The search strategy for GL is obviously different since academic databases do not index GL. The classified MLR guideline papers (as discussed in Section 3.2), identified several strategies, as discussed next:

- *General web search engine*: For example, conventional web search engines such as Google were used in many GL review studies in management [79] and health sciences [78]. This advice is valid and easily applicable in the SE context as well.
- *Specialized databases and websites*: Many papers mentioned specialized databases and websites that would be different for each discipline. For example, in medical sciences, clinical trial registries are relevant (e.g., the International Standard Randomized Controlled Trials Number, www.isrctn.com). As another example, in management sciences, investment sites have been used (e.g., www.socialfunds.com). GL database www.opengrey.eu provides broader coverage but search for "software engineering" resulted in only 4,115 hits as of this writing (March 21, 2017). For comparison, Scopus provides 120,056 hits for the same search. Relevant databases for SE would be non-peer reviewed electric archives (e.g., www.arxiv.org), social question-answer websites (e.g., www.stackoverflow.com). In essence, the choice of websites that the review authors should focus on, would depend on the particular search goals. For example, if one is interested in agile software development, a suitable website could be AgileAlliance (www.agilealliance.org). A focused source for software testing would be the website of the International Software Testing Qualifications Board (ISTQB, www.istqb.org). Additionally, many annual surveys in SE exist which provide inputs to MLRs, e.g., the World Quality Report [93], the annual state of Agile report [94], worldwide software developer and ICT-skilled worker estimates by the International Data Corporation (IDC) (www.idc.com), National-level surveys such as the survey of software companies in Finland ("*Ohjelmistoyrityskartoitus*" in Finnish) [95], or the Turkish Software Quality report [96] by the Turkish Testing Board. However, figuring out suitable specialized databases is not trivial which brings us to our next method (contacting individuals).
- *Contacting individuals directly or via social media*: Individuals can be contacted for multiple purposes for example to provide their unpublished studies or to find out specialized databases where relevant information could be searched. [79] mentions contacting individuals via multiple methods: direct requests, general requests to organizations, request to professional societies via mailing list, and open requests for information in social media (Twitter or Facebook).
- *Reference lists and backlinks*: Studying reference lists, so called snowballing [92], is done in the white (formal) literature reviews as well in GL reviews. However, in GL and in particularly GL in web sites, formal citations are often missing. Therefore, features such as backlinks can be navigated either forward or backward. Backlinks can be extracted using various online back-link checking tools, e.g., MAJESTIC (www.majestic.com).

Due to the lack of standardization of terminology in SE in general and the issue that this problem may even be more significant for the GL, the definition of key search terms in search engines and databases requires special attention. For MLRs we therefore recommend to perform an informal pre-search to find different synonyms for specific topics as well as to consult bodies of knowledge such as the Software Engineering Body of Knowledge (SWEBOK) [97] for SE in general or, for instance the standard glossary of terms used in software testing from the ISTQB [98] for testing in particular.

In MLR-AutoTest, authors used the Google search to search for GL and Google Scholar to search for academic literature. The authors used four separate search strings. In addition, forward and backward snowballing [92] was applied to include as many relevant sources as possible.

Based on the MLR goal and RQs, researchers should choose the relevant GL types and/or GL producers (data sources) for the MLR and such decisions should be made as explicit and justified as possible. Any mistake in missing certain types of GL types could lead to the final MLR output (report) missing important knowledge and evidence in the subject under study. For example, for MLR-AutoTest, we considered white papers, blog posts and even YouTube videos, and we found insightful GL resources of all these types.

---

⇒ **Guideline 6:** Identify the relevant GL types and/or GL producers (data sources) for your review study early on.

---

⇒ **Guideline 7:** General web search engines, specialized databases and websites, backlinks, and contacting individuals directly are ways to search for grey literature.

---

### 5.1.2 When to stop the search

In the formal literature, one first develops the search string and then uses this search string to collect all the relevant literature from an abstract or full text database. This brings a clear stopping condition for the search process and allows moving to study's next phases. We refer to such as a condition as data exhaustion stopping criteria. However, the issue of when to stop the GL search is not that simple. Through our own experiences in MLR studies [43, 44, 54, 62], we have observed that different stopping criteria for GL searches are needed.

First, the stopping rules are intervened with the goals and types of evidence of including GL. If evidence is mostly qualitative, one can reach theoretical saturation, i.e., a point where adding new sources do not increase the number of findings, even if one decides to stop the search before finding all the relevant sources.

Second, the stopping rules can be influenced by the large volumes of data. For example, in MLR-AutoTest, we received 1,330,000 hits from Google. Obviously, in such cases, one needs to rely on the search engine page rank algorithm [99] and choose to investigate only a suitable number of hits.

Third, stopping rules are influenced due to the varying quality and availability of evidence (see the model for differentiating the GL in Figure 1). For instance, in our review of gamification of software testing [54], the quality of evidence quickly declined when moving down in the search results provided by Google search engine. More and higher qualities for evidence were available for our MLR-AutoTest. Thus, the availability of not only resources but also the availability of evidence can determine whether data exhaustion stopping rule is appropriate.

To summarize, we offer three possible stopping criteria for GL searches:

1. Theoretical saturation, i.e., when no new concepts emerge from the search results anymore
2. Effort bounded, i.e., only include the top N search engine hits
3. Evidence exhaustion, i.e., extract all the evidence

In MLR-AutoTest, authors limited their search to the first 100 search hits and continued the search further if the hits on the last page still revealed additional relevant search results. This partially matches the "effort bounded" stopping rules augmented with an exhaustive-like subjective stopping criterion.

---

⇒ **Guideline 8:** When searching for GL on SE topics, three possible stopping criteria for GL searches are: (1) Theoretical saturation, i.e., when no new concepts emerge from the search results; (2) Effort bounded, i.e., only include the top N search engine hits, and (3) Evidence exhaustion, i.e., extract all the evidence

---

## 5.2 Source selection

Once the potentially relevant primary sources have been obtained, they need to be assessed for their actual relevance. The source selection process normally includes determining the selection criteria and performing the selection process. As GL is more diverse and less controlled than formal literature, source selection can be particularly time-consuming and difficult.

Therefore, the selection criteria should be more fine-grained and take criteria considering the source type and specific quality assessment criteria for GL, see Table 7, into account. The source selection process itself is not specific for GL, but typically more time-consuming as the selection criteria are more diverse and can be quite vague and furthermore requires a coordinated integration with the selection process for formal literature.

### 5.2.1 Inclusion and exclusion criteria for source selection

Source selection criteria are intended to identify those sources that provide direct evidence about the MLR's research (review) question. As we discuss in Section 5.3, in practice, source selection (inclusion and exclusion criteria) overlaps and is sometimes even integrated with study quality assessment [43, 44, 54, 62]. Therefore, quality assessment criteria selection, see Table 7, should also be used for the purpose of source selection. For instance, the methodology, the date of publication, or the number of backlinks can be used as a selection criterion. The benefit of this approach is that the more sources one can exclude with certainty based on suitable selection criteria, the less effort is needed for study quality assessment, which requires the more time-consuming content analysis of a source.

In MLR-AutoTest, sources were included if they are (a) in the area of automated testing ROI calculations since they could be used as a decision support mechanism for balancing and deciding between manual versus automated software testing, or (b) sources which provide decision support for the two questions "what to automate" and "when to automate". Sources that did not meet the above criteria were excluded.

⇒ **Guideline 9:** Combine inclusion and exclusion criteria for grey literature with quality assessment criteria (see Table 7).

### 5.2.2 Source selection process

The source selection process comprises the definition of inclusions and exclusion criteria, see previous section, as well as performing the process itself. The source selection process for GL requires a coordinated integration with the selection process for formal literature. Both formal and GL outlets should be investigated adequately and effort required to analyze one source type shall not reduce the effort required for the other source type. Furthermore, source selection can overlap with the searching process when searching involves snowballing or contacting the authors of relevant papers. When two or more researchers assess each paper, agreement between researchers is required and disagreements must be discussed and resolved, e.g., by voting.

In MLR-AutoTest, the same criteria were applied to GL and to formal literature. Furthermore, only inclusion criteria were provided, see Section 5.2.1 for the criteria, and sources not meeting them were excluded. The final decision on inclusion or exclusion for unclear papers was made in a voting between the two authors. The inclusions criteria were also applied in the performed forward and backward snowballing.

⇒ **Guideline 10:** In the source selection process of an MLR, one should ensure a coordinated integration of the source selection processes for grey literature and formal literature.

### 5.3 Quality assessment of sources

Quality assessment of sources is about determining the extent to which a source is valid and free of bias. Differing from formal literature, which normally follows a controlled review and publication process, processes for GL are more diverse and less controlled. Consequently, the quality of GL is more diverse and often more laborious to assess. A variety of models for quality assessment of GL sources exists, and we found the ones in [17], [50], and [70] the most well-developed.

To present a synthesized approach for quality assessment of GL sources, we used the suggestions provided in [17, 50, 51, 70] and complemented them with our own expertise in [43, 44, 54, 62] to develop the quality assessment checklist shown in Table 7. Each of our checklist criterion has strengths and weaknesses. Some are suitable only for specific types of GL sources, e.g., online comments only exist for source types open for comments like blog posts, news articles or videos. A highly commented blog post may indicate popularity, but on the other hand, spam comments may bias the number of comments, thus invalidating the high popularity.

In principle, one can use any checklist item of the quality assessment checklist for source selection as well. For instance, the methodology, the date of publication, or the number of backlinks can be used as a selection criterion. As stated before, the advantage of selection criteria is that the more sources one can exclude with certainty based on a set of criteria, the less effort is needed for the more time-consuming study quality assessment. Furthermore, when using the "research method" as the selection criterion in a specific source, e.g., survey, case study or experiment, it enables further assessment of the study quality (rigor). To investigate the quality (rigor)of specific study types in detail, checklists tailored to specific study types are available. For instance, Host and Runeson [100] presented a quality checklist for case studies, which can also be utilized for case studies reported in formal literature.

> ⇒ **Guideline 11:** Apply and adapt the criteria authority of the producer, methodology, objectivity, date, novelty, impact, as well as outlet control (e.g., see Table 7), for study quality assessment of grey literature.
>    o Consider which criteria can already be applied for source selection.
>    o There is no one-size-fits-all quality model for all types of GL. Thus, one should make suitable adjustments to the quality criteria checklist and consider reductions or extensions if focusing on particular studies such as survey, case study or experiment.

### Table 7-Quality assessment checklist of grey literature for software engineering

| Criteria | Questions |
|---|---|
| Authority of the producer | • Is the publishing organization reputable? E.g., the Software Engineering Institute (SEI) <br> • Is an individual author associated with a reputable organization? <br> • Has the author published other work in the field? <br> • Does the author have expertise in the area? (e.g. job title principal software engineer) |
| Methodology | • Does the source have a clearly stated aim? <br> • Does the source have a stated methodology? <br> • Is the source supported by authoritative, contemporary references? <br> • Are any limits clearly stated? <br> • Does the work cover a specific question? <br> • Does the work refer to a particular population or case? |
| Objectivity | • Does the work seem to be balanced in presentation? <br> • Is the statement in the sources as objective as possible? Or, is the statement a subjective opinion? <br> • Is there vested interest? E.g., a tool comparison by authors that are working for particular tool vendor <br> • Are the conclusions supported by the data? |
| Date | • Does the item have a clearly stated date? |
| Position w.r.t. related sources | • Have key related GL or formal sources been linked to / discussed? |
| Novelty | • Does it enrich or add something unique to the research? <br> • Does it strengthen or refute a current position? |
| Impact | • Normalize all the following impact metrics into a single aggregated impact metric (when data are available): Number of citations, Number of backlinks, Number of social media shares (the so-called "alt-metrics"), Number of comments posted for a specific online entries like a blog post or a video, Number of page or paper views |
| Outlet type | • 1st tier GL (measure=1): High outlet control/ High credibility: Books, magazines, theses, government reports, white papers <br> • 2nd tier GL (measure=0.5): Moderate outlet control/ Moderate credibility: Annual reports, news articles, presentations, videos, Q/A sites (such as StackOverflow), Wiki articles <br> • 3rd tier GL (measure=0): Low outlet control/ Low credibility: Blogs, emails, tweets |

The decision whether to include a source or not can go beyond a bare binary decision ("yes" or "no" informally decided on guiding question), and can be based on a richer scoring scheme. For instance, da Silva et al. [101] used a 3-point Likert scale (yes=1, partly=0.5, and no=0) to assign scores to assessment questions. Based on these scoring results, agreement between different persons can be measured and a threshold for the inclusion of sources can be defined.

As discussed above, we have not seen any of the SE MLRs (even our working example MLR-AutoTest) using such comprehensive quality assessment models. Thus, to show an example of how GL quality assessment models can be applied in practice, we apply our checklist, Table 7, to five random GL sources from the pool of MLR-AutoTest (refer to Table 8).

Table 9 shows results of applying the quality assessment checklist (in Table 7) to the five GL sources of Table 8. We provide notes for each row for justifying each assessment. The sum of the assessments and the last normalized values (each between 0-1) show the quality assessment outcome for each GL source. Out of a total quality score of 20 (the total number of individual criteria in Table 7), the five GL sources GL1, …, GL5 received the scores of 13, 19, 16, 15.5, 12, respectively. If MLR-AutoTest was conduct this type of systematic quality assessment for all the GL sources, it could for example set the quality score of 10 as the "threshold" (20/2). Any source above that would be included in the pool and any source with score below it would be excluded.

**Table 8- Five randomly-selected GL sources from candidate pool of MLR-AutoTest**

| ID | Reference |
|---|---|
| GL1 | B. Galen, "Automation Selection Criteria – Picking the "Right" Candidates," http://www.logigear.com/magazine/test-automation/automation-selection-criteria-%E2%80%93-picking-the-%E2%80%9Cright%E2%80%9D-candidates/, 2007, Last accessed: Nov. 2017. |
| GL2 | B. L. Suer, "Choosing What To Automate," http://sqgne.org/presentations/2009-10/LeSuer-Jun-2010.pdf, 2010, Last accessed: Nov. 2017 |
| GL3 | Galmont Consulting, "Determining What to Automate," http://galmont.com/wp-content/uploads/2013/11/Determining-What-to-Automate-2013_11.13.pdf, 2013, Last accessed: Nov. 2017 |
| GL4 | R. Rice, "What to Automate First," https://www.youtube.com/watch?v=eo66ouKGyVk, 2014, Last accessed: Nov. 2017 |
| GL5 | J. Andersson and K. Andersson, "Automated Software Testing in an Embedded Real-Time System," BSc Thesis. Linköping University, Sweden, 2007 |

**Table 9- Example application of the quality assessment checklist (in Table 7) to the five example GL sources (see Table 8) in the pool of MLR-AutoTest**

| Criteria | Questions | Example GL sources | | | | | Notes |
|---|---|---|---|---|---|---|---|
| | | GL1 | GL2 | GL3 | GL4 | GL5 | |
| Authority of the producer | Is the publishing organization reputable? | 0 | 0 | 1 | 0 | 0 | Only GL3 has no person as an author name, thus its authorship can only be attributed to an organization. |
| | Is an individual author associated with a reputable organization? | 1 | 1 | 0 | 1 | 1 | All sources, except GL3, are written by individual authors. |
| | Has the author published other work in the field? | 1 | 1 | 1 | 1 | 0 | GL5 is BSc thesis and a Google search for the authors' names does not return any other technical writing in this area. |
| | Does the author have expertise in the area? (e.g., job title principal software engineer) | 1 | 1 | 1 | 1 | 1 | Considered the information in the web pages |
| Methodology | Does the source have a clearly stated aim? | 1 | 1 | 1 | 1 | 1 | All five sources have a clearly stated aim. |
| | Does the source have a stated methodology? | 1 | 1 | 1 | 1 | 0 | GL5 only has a section about the topic of when and what to automate (Sec. 6.1) and that section has no stated methodology. |
| | Is the source supported by authoritative, documented references? | 0 | 1 | 0 | 0 | 1 | For GL, references are usually hyperlinks from the GL source (e.g., blog post). |
| | Are any limits clearly stated? | 0 | 1 | 1 | 1 | 0 | GL1 and GL5 are rather short and not discussing the limitations of the ideas. |
| | Does the work cover a specific question? | 1 | 1 | 1 | 1 | 1 | All five source answer the question on when and what to automate. |
| | Does the work refer to a particular population? | 1 | 1 | 1 | 1 | 1 | All five sources refer to the population of test cases that should be automated. |
| Objectivity | Does the work seem to be balanced in presentation? | 1 | 1 | 1 | 1 | 0 | We checked whether the source also explicitly looked at the issue of whether a given test case should NOT be automated. |
| | Is the statement in the sources as objective as possible? Or, is the statement a subjective opinion? | 1 | 1 | 1 | 1 | 1 | When enough evidence is provided in a source, it becomes less subjective. The original version of the question in Table 8 would get assigned '0' for the positive outcome, thus we negated it. |
| | Is there vested interest? E.g., a tool comparison by authors that are working for particular tool vendor. Are the conclusions free of bias? | 1 | 1 | 1 | 1 | 1 | The original version of the question in Table 8 would get assigned '0' for the positive outcome, thus we negated it. |
| | Are the conclusions supported by the data? | 1 | 1 | 1 | 1 | 1 | |
| Date | Does the item have a clearly stated date? | 0 | 1 | 1 | 1 | 1 | GL1 is a website and does not have a clearly stated date related to its content. |
| Position w.r.t. related sources | Have key related GL or formal sources been linked to / discussed? | 0 | 1 | 0 | 0 | 1 | For GL sources, references (bibliography) are usually the hyperlinks from the source (e.g., blog post). |
| Novelty | Does it enrich or add something unique to the research? | 1 | 1 | 1 | 1 | 0 | GL5 does not add any novel contribution in this area. Its focus is on test automation, but not on the when/what questions. |
| | Does it strengthen or refute a current position? | 1 | 1 | 1 | 1 | 0 | GL5 does not add any novel contribution in this area. Its focus is on test automation, but not on the when/what questions. |
| Impact | Normalize all the following impact metrics into a single aggregated impact metric (when data are available): Number of citations, Number of backlinks, Number of social media shares (the so-called "alt-metrics"), Number of comments posted for a specific online entries like a blog post or a video, Number of page or paper views | 0 | 1 | 0 | 0 | 0 | -For backlinks count, we used this online tool: http://www.seoreviewtools.com/valuable-backlinks-checker/ -GL3 became a broken link at the time of this analysis. -Only GL2 had two backlinks. The others had 0. -All other metrics values were 0 for all five sources. -For counts of social media shares, we used www.sharedcount.com |
| Outlet type | • 1st tier GL (measure=1): High outlet control/ High credibility: Books, magazines, theses, government reports, white papers<br>• 2nd tier GL (measure=0.5): Moderate outlet control/ Moderate credibility: Annual reports, news articles, videos, Q/A sites (such as StackOverflow), Wiki articles<br>• 3rd tier GL (measure=0): Low outlet control/ Low credibility: Blog posts, presentations, emails, tweets | 0 | 1 | 1 | 0.5 | 1 | GL1: blog post GL2, GL3: white papers GL4: YouTube video GL5: thesis |
| | Sum (out of 20): | 13 | 19 | 16 | 15.5 | 12 | Summation of the values in the previous rows |
| | Normalized (0-1): | 0.65 | 0.95 | 0.80 | 0.78 | 0.60 | Normalized values by dividing the values in the previous row by 20 (number of factors) |

## 5.4 Data extraction

For the data extraction phase, we discuss the following aspects:

- Design of data extraction forms
- Data extraction procedures and logistics
- Possibility of automated data extraction and synthesis

### 5.4.1 Design of data extraction forms

Most of design aspects of data extraction forms for MLRs are similar to those aspects in the SLR guidelines of Kitchenham and Charters [22]. We discuss next a few important additional considerations in the context of MLRs based on our experience in conducting MLRs.

Since many SLR and MLR studies have, as a part of them, an SM step first, it is important to ensure the rigor in data extraction forms. Also, in a recent paper [102] we presented a set of experience-based guidelines for effective and efficient data extraction which can apply to all four types systematic reviews in SE (SM, SLR, MLR and GLRs).

Based on the suggestions in [102], to facilitate design of data extraction forms, we have developed spreadsheets with direct traceability to MLR's research questions in mind. For example, Table 10 shows the systematic map that we used developed and used in the MLR-AutoTest. In this table, column 1 is the list of RQs, column 2 is the corresponding attribute/aspect. Column 3 is the set of all possible values for the attribute. Finally, column 4 indicates for an attribute whether multiple selections can be applied. For example, in RQ 1 (research type), the corresponding value in the last column is 'S' (Single). It indicates that one source can be classified under only one research type. In contrast, for RQ 1 (contribution type), the corresponding value in the last column is 'M' (Multiple). It indicates that one study can contribute more than one type of options (e.g., method, tool, etc.).

According to our experience and due to the issue that GL sources have a less standardized structure than formal literature, it is also useful to provide "traceability" links (i.e., comments) in the data extraction form to indicate the position in the GL source where the extracted information was found. This issue is revisited again in the next subsection (see Figure 8).

#### Table 10: Systematic map developed and used in MLR-AutoTest

| RQ | Attribute/Aspect | Categories | (M)ultiple/ (S)ingle |
|---|---|---|---|
| - | Source type | Formal literature, GL | S |
| 1 | Contribution type | Heuristics/guideline, method (technique), tool, metric, model, process, empirical results only, other | M |
| | Research type | Solution proposal, validation research (weak empirical study), evaluation research (strong empirical study), experience studies, philosophical studies, opinion studies, other | S |
| 2 | Factors considered for deciding when/what to automate | A list of pre-defined categories (Maturity of SUT, Stability of test cases, 'Cost, benefit, ROI', and Need for regression testing) and an 'other' category whose values were later qualitatively coded (by applying 'axial' 'open' coding) | M |
| 3 | Decision-support tools | Name and features | M |
| 4 | Attributes of the software systems under test (SUT) | Number of software systems: integer SUT names: array of strings Domain, e.g., embedded systems Type of system(s): Academic experimental or simple code examples, real open-source, commercial Test automation cost/benefit measurements: numerical values | M |

### 5.4.2 Data extraction procedures and logistics

Many authors are reporting logistical and operational challenges in conducting SLRs, e.g., [63]. We suggest next a summary of best practices based on our survey of MLR guidelines - 4 out of 24 MLR guideline papers, see Section 3.2, provided experience-based advices for data extraction - as well as our own experiences.

Authors in [25] offered a worksheet sample to extract data from the GL sources including fields such as: database, organization, website, pathfinder, guide to topic/subject, date searched, # of hits, and observations.

In [79] the authors emailed and even called individuals to gather more detailed GL data. For GL, often only a subset of the original important data is made available in the GL source (to keep it short and brief) and detailed information is only available in "peoples' heads" [79].

The authors of [80] found cases where both the grey and peer-reviewed documents described the same study. In those cases, the team decided the primary document would be the peer-reviewed with GL documents as supplemental.

The guidelines in [12] suggested maintaining "*chains of evidence (records of sources consulted and inferences drawn)*". This is similar to what we call "traceability" links in SE as highlighted before and also suggest in our previous data extraction guidelines [102]. For instance, Figure 8 shows a snapshot of the online repository (spreadsheet hosted on Google Docs) for MLR-AutoTest in which the contribution facets are shown and a classification of 'Model' for the contribution facet of [Source 2] is shown, along with the summary text from the source acting as the corresponding traceability link.

When traceability information (verbatim text from inside the primary studies) are not included in the data extraction sheets, peer reviewing of the data by other team members, and also finding the exact locations in the primary studies where the data actually come from become challenging. We have experienced such a challenge in many occasions in our past MLR and SLRs.

Furthermore, the authors of [12] also argue that, because documents in the GL are often written for non-academic purposes and audiences and because documents often address different aspects of a phenomenon with different degrees of thoroughness, it is essential that researchers record the purpose and specify the coverage of each GL document. These items are also in our quality assessment checklist in Table 7. Authors of [12] also wanted to record the extent to which the implicit assumptions or causal connections were supported by evidence in the GL documents. For that purpose, they developed matrices that enabled them to systematically track and see what every document in every category of the dataset said about causal connection in every theory of action.

Our own experiences from our past SLRs and MLRs have been as follows. To extract data, the studies in our pool were reviewed with the focus of each specific RQ. Researchers should also extract and record as much quantitative/qualitative data as needed to sufficiently address each RQ. If not, answering the RQ under study will be impossible based on inadequate extracted data and would require further efforts to review, read and extract the missing data from the primary studies again. We have experienced such a challenge in many occasions in our past MLR and SLRs. During the analysis, each involved researcher extracted and analyzed data from the share of sources assigned to her/him, then each researcher peer reviewed the results of each other's analyses. In the case of disagreements, discussions were conducted. We utilized this process to ensure quality and validity of our results.

**Figure 8: A snapshot of the publicly-available spreadsheet hosted on Google Docs for MLR-AutoTest. The full final repository can be found in http://goo.gl/zwY1sj.**

⇒ **Guideline 12:** During the data extraction, systematic procedures and logistics, e.g., explicit "traceability" links between the extracted data and primary sources, should be utilized. Also, researchers should extract and record as much quantitative/qualitative data as needed to sufficiently address each RQ, to be used in the synthesis phase.

## 5.5 Data synthesis

There are various data synthesis techniques, as reported in Kitchenham and Charters' guidelines for SLRs [22] and elsewhere. For instance, a guideline paper for synthesizing evidence in SE research [103] distinguishes descriptive (narrative) synthesis, quantitative synthesis, qualitative synthesis, thematic analysis, and meta-analysis.

Based on the type of RQs and the type of data (primary studies), the right data synthesis techniques should be selected and used. We have observed that practitioners provide in their reports mainly three types of data:

- First, qualitative and experience-based evidence is very common in the GL as practitioners share their reflections on topics such as on when to automate testing (MLR-AutoTest). This requires qualitative data analysis techniques. Their reflection may occasionally include quantitative data, e.g., some presented quantitative data on ROI when automating software testing. However, we see that typically the quality and accuracy of the reporting does not allow to conduct quantitative meta-analysis from practitioner GL reports.

- Second, quantitative evidence in the form of questionnaires is relatively common in GL, e.g., international surveys such as the state-of the Agile report by VersionOne, and the World Quality Survey by HP & Sogetti. More surveys can be found on national/regional levels such as the survey of software companies in Finland [95], or the Turkish Software Quality report [96] by the Turkish Testing Board. If the same questionnaire is repeated in multiple or sequential surveys, this may allow meta-analysis. However, often the GL surveys fail to report standard deviation, which makes statistical meta-analysis impossible. Furthermore, we have seen virtually no controlled experiments or rigorously conducted quasi-experiments in GL, thus, we see limited possibilities in using meta-analytic procedures to combine experiment results from GL in SE.

- Third, using data from particular GL databases such as question/answer sites (such as the StackOverflow website) may allow both the use of quantitative and qualitative research methods. For example, a quantitative comparison of technology usage can be done from the StackOverflow website by extracting the number of questions and view counts that can give an indication of popularity of testing tools for example [104]. Qualitative analysis (such as open coding and grounded theory) [105] can also be conducted and it can analyze the types of problems software engineers are facing with the testing tools for example.

In MLR-AutoTest, authors conducted qualitative coding [105] to derive the factors for deciding when to automate testing. We had some pre-defined factors (based on our past knowledge of the area), namely "regression testing", "maturity of SUT" and "ROI". During the process, we found out that our pre-determined list of factors was greatly limiting, thus, the rest of the factors emerged from the sources, by conducting "open" and "axial coding" [105]. The creation of the new factors in the "coding" phase was an iterative and interactive process in which both researchers participated. Basically, we first collected all the factors affecting when- and what-to-automate questions from the sources. Then we aimed at finding factors that would accurately represent all the extracted items but at the same time not be too detailed so that it would still provide a useful overview, i.e., we chose the most suitable level of "abstraction" as recommended by qualitative data analysis guidelines [105].

Figure 9 shows the phases of qualitative data extraction for the factors, where the process started from a list of pre-defined categories: stability (maturity) of SUT, stability of test cases, 'cost, benefit, ROI, and need for regression testing) and a large number of raw factors phrased under the "Other" category. By an iterative process, those phrases were qualitatively coded (by applying axial and open coding approaches [105]) to yield the final result, i.e., a set of cohesive well-grouped factors.

(a): initial phase of qualitative coding

(b): final result of qualitative coding

**Figure 9: Phases of qualitative data extraction for factors considered for deciding when/what to automate (taken from the MLR-AutoTest paper)**

For data synthesis from GL sources, utilizing the argumentation theory also can be useful. As discussed in Section 2.3, there has been some recent work in SE on extracting SE practitioners' evidence and beliefs, e.g., the study by Rainer [51]. One of the interesting materials presented in [51] was a set of critical questions for using argumentation from expert opinions in GL, as follows:

1. Expertise: How credible is W ("Writer" of the GL source) as an expert source?

2. Field: Is W an expert in the field that P is in?
3. Opinion: What did W assert that implies P?
4. Trustworthiness: Is W personally reliable as a source?
5. Consistency: Is proposition P consistent with what other experts assert?
6. Backup evidence: Is W's assertion based on evidence?

In the above questions, *P* is a proposition in a GL source, and *W* refers to the writer of a GL source, e.g., a SE practitioner who writes her/his opinion in a blog. The questions were adopted from a book on argumentation theory [52]. While the above questions seem to be useful and rationale, some of them seem slightly questionable, e.g., question #4 cannot be reliably assessed. Also question #5 could be irrelevant since experts should be allowed to have non-consistent opinions with each other.

Furthermore, researchers should carefully balance synthesis using sources with different levels of rigor. We can easily see that the rigor used in a blog post is different than that of a research paper, and when synthesizing evidence from both types, their contributions to the combined evidence would ideally be not in the same "amount" (weight). Earlier in Section 5.3, we discussed checklists for quality assessment of formal and GL sources: using the checklist presented by Host and Runeson [100] for case studies reported in formal literature, and the checklist in Table 7 for GL sources. By carefully combining the two chosen checklists, we may be able to objective assign evidence (rigor) weights to different sources and thus synthesize evidence from all types in a more systematic manner.

⇒ **Guideline 13:** A suitable data synthesis method should be selected. Many GL sources are suitable for qualitative coding and synthesis. Some GL sources allow combination of survey results but lack of reporting rigor limits the meta-analysis. Quantitative analysis is possible on GL databases such as StackOverflow. Also argumentation theory can be beneficial for data synthesis from grey literature. Finally, the limitations of GL sources w.r.t. their evidence depth of experiment prevent meta-analysis.

## 6 REPORTING THE REVIEW

As shown in the MLR process, see Figure 7, the last phase is reporting the review. Typical issues of the reporting phase of an MLR are similar to the SLR guidelines of Kitchenham and Charters [22]. In the experience from our past SLR and MLRs, we have seen two important additional issues that we discuss next: (1) reporting style for different audience types, and (2) ensuring usefulness to the target audience.

MLR needs to provide benefits for both researchers and practitioners since it contains a summary of both the state-of-the-art and –practice in a given area. Readers of MLR papers (both practitioners and researchers) are expected to benefit from the evidence-based overview and index to the body of knowledge in the given subject area [62].

Furthermore, conveying and publishing results of MLR and SLR studies to practitioners will "*enhance the practical relevance of research*" [106]. To enhance the practical relevance of research, [106] suggested to "*convey relevant insights to practitioners*", "*present to practitioners*" and "*write for practitioners*". We have followed that advice, and have reported the shortened (practitioner-oriented) versions of three of our MLR and SLR studies [9, 62, 107] in the IEEE Software magazine. When the reporting style of SLR or an MLR is "fit" for practitioners, they usually find such papers useful.

We have also found it useful to ask for practitioners' feedback to make the results even more communicative. We recommend including in review papers a section about the implications of the results, as we reported in [43, 44], and if possible, a section on the benefits of the review. For example, in our SLR on testing embedded software [9], we included a section on "*Benefits of this review*". To further assess the benefits of our review study in [9], we asked several active test engineers in the Turkish embedded software industry to review the review paper and the online spreadsheet of papers, and let us know what they think about the potential benefits of that review paper. Their general opinion was that a review paper like that paper [9] is a valuable resource and can actually serve as an index to the body of knowledge in this area.

Furthermore, reporting style for scientific journals and practitioners' magazines are quite different [106, 108]. While papers in scientific journals should provide all the details of the MLR (the planning and search process), papers in practitioner-oriented outlets (such as IEEE Software) should be in general shorter, succinct and "to the point". We have been aware of this issue and have followed slightly different reporting styles in the set of our recent MLRs/SLRs. For example, we wrote [62, 107] and published them in the IEEE Software magazine targeting practitioners, while we wrote their extended scientific versions afterwards and published in the Information and Software Technology journal [5, 43]. Table 11 shows our publication strategy for three sets of recent MLR/SLR studies, on three testing-related topics: test maturity and test process improvement, testing embedded software, and software test-code engineering.

With respect to MLR-AutoTest, we did not have an IEEE Software publication, however, our academic paper in Information and Software Technology includes a practitioner oriented list questions that can be used in deciding whether to automate testing or not in particular context. An excerpt of that checklist, as taken from MLR-AutoTest, is shown in Table 12. Thus, a practitioner oriented section even if the authors do not wish to make to separate publications.

**Table 11- Publication strategy of three sets of MLR/SLR studies**

| MLR/SLR topic | Paper title | Ref. | Journal/magazine | Main audience | |
|---|---|---|---|---|---|
| | | | | Researchers | Practitioners |
| Test maturity and test process improvement | What we know about software test maturity and test process improvement | [62] | IEEE Software | | x |
| | Software test maturity assessment and test process improvement: a multivocal literature review | [43] | Information and Software Technology | x | |
| Testing embedded software | What we know about testing embedded software | [9] | IEEE Software | | x |
| | Testing embedded software: a systematic literature review | | Submitted to a journal | x | |
| Software test-code engineering | Developing, verifying and maintaining high-quality automated test scripts | [107] | IEEE Software | | x |
| | Software test-code engineering: a systematic mapping | [5] | Information and Software Technology | x | |

**Table 12- Excerpt of a practitioner-oriented checklist of whether to automate testing or not (taken from MLR-AutoTest)**

| Category | Area (weight, i.e., num. of sources) | Situation | +/- |
|---|---|---|---|
| SUT-related factors | Maturity of SUT (39) | SUT or the targeted components will experience major modifications in the future. | - |
| | | The interface through which the tests are conducted is unlikely to change. | + |
| | Other SUT aspects (6) | SUT is an application with a long life cycle. | + |
| | | SUT is a generic system, i.e. not tailor made or heavily customized system. | + |
| | | SUT is tightly integrated into other products, i.e. not independent. | - |
| | | SUT is complex. | - |
| | | SUT is mission critical. | + |

Another issue is choosing suitable and attractive titles for papers targeting practitioners [109]. In two of our review papers published in IEEE Software [9, 62], we entitled them starting with "*What we know about …*". This title pattern seems to be attractive to practitioners, and has also been used by other authors of IEEE Software papers [110-114]. Another IEEE Software paper [115] showed, by textual analysis, that practitioners usually prefer simpler phrases for the titles of their talks at conferences or their (grey literature) reports, compared with more complex titles used in the formal literature.

A useful resource that the authors of MLR/SLR should publish as a public online version is the repository of the review studies included in the MLR which many researchers will find a useful add-on to the MLR itself. Ideally, the online repository comes with additional export, search and filter functions to support further processing of the data. Figure 8 from Section 5.4 shows an example of an online paper repository implemented as a Google Docs spreadsheet, i.e., the list of included sources of MLR-AutoTest. Such repositories provide various benefits, e.g., transparency on the full dataset, replication and repeatability of the review, support when updating the study in the future by the same or a different team of researchers, and easy access to the full "index" of sources.

One of the earliest online repositories serving as companion to its corresponding survey papers [116, 117] and showing the usefulness of such online repositories, is the one on the subject of Search-based software Engineering (SBSE) [118], which was first published in 2009 and has been actively maintained since then.

⇒ **Guideline 14:** The writing style of an MLR paper should match its target audience, i.e., researchers and/or practitioners.

> - o If targeting practitioners, a plain and to-the-point writing style with clear suggestion and without details about the research methodology should be chosen. Asking feedback from practitioners is highly recommended.
> - o If the MLR paper targets researchers, it should be transparent by covering the underlying research methodology as well as an online repository and highlight the research findings while providing directions to future work.

## 7 CONCLUSIONS AND FUTURE WORKS

We think that software engineering research can improve its relevance by accepting and analyzing input from practitioner literature. Currently, books and consultancy reports are considered valid evidence while relevant input found in blogs and in social media discussions is often ignored. Furthermore, practitioner interviews done and reported by researchers have, for long, been considered as academic evidence in empirical software engineering, while grey literature produced by the very same individuals may have been ignored as unscientific. This paper wants to lift such a double standard by allowing rigorously conducted analysis of practitioners' writings to enter the scientific literature.

As existing guidelines for performing systematic literature studies in SE provide limited coverage for including the practitioners sources and conducting multivocal literature reviews, this paper filled this gap by developing and presenting a set of experience-based guidelines for planning, conducting and presenting MLR studies in SE.

To develop the MLR guidelines, we benefited from three inputs: (1) existing SLR and SM guidelines in SE, (2) a survey of MLR guidelines and experience papers in other fields, and (3) our own experiences in conducting several MLRs in SE. We took the popular SLR guidelines of Kitchenham and Charters as the baseline and extended it to conduct MLR studies. The presented guidelines covered all phases of conducting and reporting MLRs in SE from the planning phase, to conducting the review, and to reporting the review. In particular, we believe that incorporating and adopting a vast set of experience-based recommendations from MLR guidelines and experience papers in other fields enabled us to propose a set of guidelines with solid foundations.

We should also note the limitations of the guidelines that we have developed and presented in this paper: (1) although they are based on our previous experience and the guidelines in other fields, they still need to be empirically evaluated in future studies; and (2) similar to any set of guidelines, our guidelines are based on our experience and also synthesis of other studies, and thus personal researcher bias could be involved, but we have mitigated such bias to the best of our ability.

The authors recommend the researchers to apply the guidelines in conducting MLR studies, and then share their lessons learned and experiences with the community. Guidelines like the ones reported in this paper are *living* entities, and have to be assessed and improved in several iterations.

We suggest future works in the following directions. First, based on guidelines such as [119] in the educational sciences field, we suggest specific guidelines and considerations for different types of reviews: systematic review, best-evidence synthesis, narrative review and for different objectives: integrative research review, theoretical review, methodological review, thematic review, state-of-the-art review, historical review, comparison of two perspectives review and review complement. Second, improving the guidelines based experiences in applying them. Third, refine guidelines for specific types of grey literature sources like blog articles or specific SE areas.

## ACKNOWLEDGMENTS

The third author has been partially supported by the Academy of Finland Grant no 298020 (Auto-Time) and by TEKES Grant no 3192/31/2017 (ITEA3: 16032 TESTOMAT project).